\documentclass[11pt]{article}
\usepackage[margin=1in]{geometry}
\usepackage{fontspec,xeCJK}
\setCJKmonofont[Path=fonts/]{cwTeXQMingZH-Medium.ttf}
\usepackage[T1]{fontenc}
\usepackage{microtype}
\usepackage{amsmath,amssymb,amsthm,mathtools}
\usepackage[sc,osf]{mathpazo}
\usepackage{booktabs,graphicx,array}
\usepackage[round,authoryear]{natbib}
\usepackage[hidelinks]{hyperref}
\numberwithin{equation}{section}
\newcommand{\E}{\mathbb E}
\newcommand{\Prob}{\mathbb P}
\newcommand{\Var}{\operatorname{Var}}
\newcommand{\Cov}{\operatorname{Cov}}

\DeclareMathOperator*{\argmin}{arg\,min}
\DeclareMathOperator{\diag}{diag}
\hypersetup{pdftitle={Recovering Unobserved Network Links from Aggregated Relational Data},pdfauthor={Yen-hsuan Tseng}}
\title{Recovering Unobserved Network Links from Aggregated Relational Data:\\
Bayesian Latent Surface Modeling and Penalized Regression}
\author{Yen-hsuan Tseng}
\date{September 2026}
\begin{document}
\maketitle
\begin{center}\small
Revised working manuscript. Based on the August 2025 master's thesis.\\
This revision is separate from the approved university version.
\end{center}
\begin{abstract}
Aggregated relational data (ARD) record counts of ties to attribute-defined groups while leaving individual edges unobserved. We compare latent-geometry and regularized network estimators through a common observation map. The comparison distinguishes the realized adjacency matrix, conditional edge probabilities, and model parameters. We study roster-based ARD with known node-level group memberships, giving both estimators the same roster and aggregate counts.

We relate the aggregate means to a Poisson working likelihood and a Huber loss, and give their derivatives. Overlapping groups, shared edges, and reporting error affect the interpretation of these objectives. Geometry restricts the representation of edge probabilities, while regularization selects among candidate fits. Identification depends on the observation map and model restrictions rather than uniqueness of a numerical optimizer.

A reproducible synthetic experiment specifies the data-generating process, estimation algorithms, and evaluation targets under matched information. The matrix estimator gives better realized-edge rankings and aggregate fit, while the geometric estimator gives lower error for generating probabilities. The resulting framework organizes ARD reconstruction around the interaction of observation design, structural assumptions, and computation.
\end{abstract}


\section{Introduction}\label{chap:intro}
\subsection{The reconstruction problem}
Network data describe relationships among individuals, institutions, or other units. Collecting a complete adjacency matrix can be expensive or incompatible with privacy restrictions. ARD replace individual neighbor lists with counts: a respondent reports how many contacts belong to specified groups. The groups may be defined by demographic characteristics, occupations, or institutional categories.

The data pose an inverse problem. A count identifies neither the members of the reported neighbor set nor the links among them. Reconstruction combines these counts with restrictions on the network or its edge probabilities. Latent space models and penalized estimators impose those restrictions differently.

We compare a latent surface model with a regularized probability-matrix estimator. The former uses a geometric representation of nodes; the latter fits the aggregate observations directly under matrix constraints. A shared observation design makes clear which differences arise from the estimators rather than their inputs.

\subsection{The observation regime}
Our setting includes a roster of all \(n\) nodes and their group memberships. Only edges are unobserved. This setting occurs when administrative covariates or an institutional roster are available alongside aggregate relationship counts. It is more informative than a survey that observes only ARD and group population totals.

When individual memberships are unknown, a likelihood that explicitly sums over the members of each group cannot be evaluated as written. That survey setting requires a model integrating over the unobserved memberships or latent group distributions, as in the foundational latent-surface literature \citep{mccormick2015,breza2020}. Here the roster is part of the observed data, not hidden information supplied to one estimator.

\subsection{Targets and organization}
We distinguish realized adjacency, conditional edge probabilities, and structural parameters. A fitted probability matrix is not itself a realized network. A sampled or thresholded adjacency is a derived output with its own interpretation.

We first formulate the observation map and estimators, then examine identification and report a matched-input experiment. Appendix~\ref{app:derivatives} gives the likelihood and loss derivatives.

\section{Observation Design and Related Methods}\label{chap:observation}
\subsection{From edges to group counts}
Let \(A=(a_{ij})\) be an undirected adjacency matrix with \(a_{ii}=0\). Let \(B\in\{0,1\}^{n\times K}\) record group memberships: \(B_{jk}=1\) when \(j\in G_k\). For respondent \(i\),
\begin{equation}
y_{ik}=\sum_{j\ne i}B_{jk}a_{ij}.\label{eq:ard}
\end{equation}
Groups may overlap in the general formulation. If all nodes respond, \(Y=AB\); if only nodes in \(\mathcal R\) respond, \(Y_{\mathcal R}=A_{\mathcal R,:}B\). The experiment uses a balanced partition and all respondent rows.

Let \(p_{ij}(\theta)=\Prob(a_{ij}=1\mid\theta,B)\). With \(B\) treated as fixed,
\begin{equation}
\mu_{ik}(\theta)=\E(y_{ik}\mid\theta,B)
=\sum_{j\ne i}B_{jk}p_{ij}(\theta).\label{eq:mean}
\end{equation}
This equality uses linearity of expectation. The identity \(Y=AB\) concerns realized edges and counts; equation~\eqref{eq:mean} concerns their conditional expectations.

\subsection{Dependence in aggregate counts}
Under conditional independence of distinct undirected edges, each count is Poisson-binomial:
\begin{equation}
\Var(y_{ik}\mid\theta,B)=\sum_{j\ne i}B_{jk}p_{ij}(1-p_{ij}).
\end{equation}
Overlapping groups share edges within a row:
\begin{equation}
\Cov(y_{ik},y_{i\ell}\mid\theta,B)
=\sum_{j\ne i}B_{jk}B_{j\ell}p_{ij}(1-p_{ij}).
\end{equation}
For \(i\ne j\), two respondent rows share the edge \(a_{ij}\), giving
\begin{equation}
\Cov(y_{ik},y_{j\ell}\mid\theta,B)
=B_{jk}B_{i\ell}p_{ij}(1-p_{ij})
\end{equation}
under the same conditional independence assumption. These expressions explain why a product of marginal count likelihoods can be a working objective rather than the true joint likelihood.

\subsection{Foundational models}
Latent space models express tie probabilities through node positions \citep{hoff2002}. \citet{mccormick2015} develop latent surface models for networks using ARD, and \citet{breza2020} estimate network-formation parameters and use the fitted model to infer network statistics without complete network data. \citet{breza2023} analyze consistent estimation of network statistics from ARD. Their assumptions, observation regimes, and target statistics are integral to their results.

\citet{alidaee2020} estimate link probabilities using penalized regression. Their matrix-based approach regularizes the probability matrix. The penalty, parameterization, and observation restrictions jointly define the estimator.

Our comparison uses a logistic geometric working model and a direct probability-matrix comparator. These are explicit implementations for the roster-based setting, not claimed reproductions of every procedure in the cited papers.

\section{Latent Geometry and Bayesian Computation}\label{chap:bayes}
\subsection{Geometric representation}
The general logistic spherical specification is
\begin{equation}
p_{ij}(\theta)=\sigma(\nu_i+\nu_j+\zeta z_i^\top z_j),
\quad z_i\in\mathbb S^p,\quad
\sigma(t)=\frac{1}{1+\exp(-t)},\quad\zeta>0.
\label{eq:edge}
\end{equation}
Intercepts describe link propensities and inner products describe geometric similarity. This is the logistic variant retained from the original manuscript; results for other link functions require an explicit correspondence.

The aggregate mean is obtained by summing~\eqref{eq:edge} over known group members. The empirical comparison uses the circular specialization \(z_i=(\cos t_i,\sin t_i)\), a common intercept, and fixed geometric strength. This removes node-specific intercept estimation from the experiment and makes its parameterization explicit.

\subsection{Poisson working likelihood}
The working objective is
\begin{equation}
\ell_W(\theta)=\sum_{i\in\mathcal R,k}
\{y_{ik}\log\mu_{ik}(\theta)-\mu_{ik}(\theta)-\log(y_{ik}!)\}.
\label{eq:working}
\end{equation}
A Poisson approximation to a Bernoulli sum requires rare-event behavior; large group size alone is insufficient when individual probabilities remain substantial. We use~\eqref{eq:working} as a working likelihood, including in the finite simulation where it need not closely approximate every marginal distribution.

For positive means,
\begin{equation}
\nabla\ell_W(\theta)=\sum_{i\in\mathcal R,k}
\frac{y_{ik}-\mu_{ik}(\theta)}{\mu_{ik}(\theta)}
\nabla\mu_{ik}(\theta).
\label{eq:score}
\end{equation}
Equivalently, the summand is \((y_{ik}-\mu_{ik})\nabla\log\mu_{ik}\). The inverse-mean factor is essential.

\subsection{Posterior target and computation}
A prior \(\pi(\theta)\) yields the working posterior
\begin{equation}
\pi_W(\theta\mid Y,B)\propto \exp\{\ell_W(\theta)\}\pi(\theta).
\end{equation}
For general spherical models, candidate priors include normal node intercepts and uniform or von Mises--Fisher latent positions. The numerical experiment fixes a smaller parameterization and fully specifies its prior and sampler in Section~\ref{chap:numerical}.

A posterior mean of edge probabilities averages over retained parameter draws. It differs from evaluating probabilities at a posterior mean of latent coordinates. Coordinate averaging is especially problematic when different draws have different orientations.

\subsection{Invariance and uncertainty}
Replacing every \(z_i\) by \(Qz_i\), for orthogonal \(Q\), preserves all edge probabilities. Alignment can be used for coordinate comparisons; probability matrices themselves are invariant. Removing an arbitrary orientation does not establish identification of all remaining parameters.

Working-posterior uncertainty reflects the likelihood and prior used. Dependence omitted from the working likelihood affects calibration. The experiment evaluates point reconstruction and computational diagnostics; it does not turn posterior variability into a validated frequentist coverage statement.

\section{Regularized Estimation and Robust Losses}\label{chap:penalized}
\subsection{Direct probability estimation}
A matrix formulation estimates \(P=(p_{ij})\) under symmetry, zero diagonal, and \(0\le p_{ij}\le1\). Its predicted counts are \(PB\). A regularized objective takes the form
\begin{equation}
\widehat P\in\argmin_{P\in\mathcal C}
\{L(Y,PB)+\lambda\mathcal P(P)\},
\quad
\mathcal C=\{P=P^\top,\ \diag P=0,\ 0\le P\le1\}.
\label{eq:matrix}
\end{equation}
This uses the same \(Y\) and \(B\) as the geometric estimator. A matrix penalty controls a property of \(P\), rather than introducing unavailable pairwise covariates.

An alternative regression formulation sets \(p_{ij}=\sigma(x_{ij}^\top\beta)\) and penalizes \(\beta\). That procedure needs observed \(x_{ij}\). It is therefore not the no-covariate comparator in the present experiment.

\subsection{What the penalty does}
An \(\ell_1\) penalty applied directly to dyadic logits \(\operatorname{logit}(p_{ij})\) shrinks them toward zero, which corresponds to probability \(1/2\). This differs from penalizing regression coefficients in a model with an intercept or other offsets, and does not directly promote sparse edges. Matrix penalties must likewise be interpreted on their actual domain: nuclear-norm and quadratic penalties impose different forms of regularization.

A unique optimizer selects one fit according to the loss and penalty. It does not establish that the data identify a unique network. Regularization and identification are separate questions.

\subsection{Reporting error}
With perturbed counts \(\widetilde y_{ik}=y_{ik}+\epsilon_{ik}\), additive noise may create negative or noninteger observations. A residual loss can replace a literal Poisson likelihood:
\begin{equation}
Q_n(\theta)=\sum_{i\in\mathcal R,k}
\rho_\xi(\widetilde y_{ik}-\mu_{ik}(\theta)),
\quad
\rho_\xi(u)=
\begin{cases}u^2/2,&|u|\le\xi,\\
\xi|u|-\xi^2/2,&|u|>\xi.\end{cases}
\label{eq:huber}
\end{equation}
Its derivative \(\psi_\xi\) clips the residual at \(\pm\xi\). Writing \(r_{ik}=\widetilde y_{ik}-\mu_{ik}(\theta)\),
\begin{equation}
\nabla Q_n(\theta)=-\sum_{i\in\mathcal R,k}
\psi_\xi(r_{ik})\nabla\mu_{ik}(\theta).
\label{eq:huber-score}
\end{equation}
The minus sign follows from differentiating the residual.

\subsection{The robust population target}
Let \(Q(\theta)=\E Q_n(\theta)\). Mean-zero residuals do not generally imply \(\E\psi_\xi(r_{ik})=0\), nor do they imply \(\E\rho_\xi(r_{ik})=0\). At an interior stationary population target \(\theta_\star\),
\begin{equation}
\sum_{i\in\mathcal R,k}
\E\{\psi_\xi(\widetilde y_{ik}-\mu_{ik}(\theta_\star))\}
\nabla\mu_{ik}(\theta_\star)=0,
\end{equation}
when differentiation and expectation can be interchanged. Agreement between this target and a mean-model parameter must be established from the observation model. The loss definition alone does not supply consistency.

\section{Identification and Local Estimation}\label{chap:identification}
\subsection{The aggregate-mean map}
For fixed \(B\) and respondents, let \(M(\theta)=(\mu_{ik}(\theta))\). Under an independent Poisson working model, equal values of \(M\) produce the same working count distribution. Identification asks whether that distribution determines the target after accounting for invariances.

Scale-free degree distributions and Watts--Strogatz graph constructions do not, by themselves, specify how traits distinguish parameters of a latent surface model. Combining these descriptions requires an explicit joint model. Similarly, a result for a maximum-likelihood estimator does not automatically characterize a posterior mean or penalized fit.

For an exact dependent ARD model, the full law may carry information beyond its mean. A conclusion concerning \(M\) must be interpreted for the mean-based or working model under consideration.

\subsection{Local sensitivity}
Let \(\eta\) be local coordinates after removing redundant orientation, and \(J\) the Jacobian of \(M\). Expected negative curvature of the Poisson objective is
\begin{equation}
H(\eta)=J(\eta)^\top\diag\{1/\mu_{ik}(\eta)\}J(\eta).
\end{equation}
Local information requires changes in target coordinates to affect aggregate means. Bounds on individual derivative norms do not bound every direction; directional sensitivity concerns singular values of the weighted Jacobian.

Local curvature and global uniqueness remain distinct. Sensitivity near one parameter does not exclude a distant observationally equivalent parameter. A growing parameter dimension also requires its own asymptotic argument rather than an unqualified appeal to a fixed-dimensional law of large numbers.

\subsection{Local score expansion}
A score expansion relates the estimation error to the score and inverse curvature. For example, an inverse-curvature bound of order \((\log n)/n\) and a score bound of order \(O_P(\sqrt{n\log n})\) yield the product
\begin{equation}
\frac{\log n}{n}\,O_P(\sqrt{n\log n})
=O_P\!\left(\frac{(\log n)^{3/2}}{\sqrt n}\right).
\end{equation}
The two bounds require score concentration and directional curvature control for the estimator, observation dependence, and parameter sequence.

In a fixed-dimensional regular M-estimation regime, a first-order score expansion gives the covariance approximation \(H^{-1}VH^{-1}\), where \(H=-\E\nabla^2\ell_W\) and \(V=\Var(\nabla\ell_W)\) both refer to the total working objective. Equality \(V=H\) depends on the joint model. Growing-dimensional latent parameters require a separate expansion with dimension-dependent remainder control.

\section{Matched-Input Synthetic Experiment}\label{chap:numerical}
\subsection{Data generation and shared inputs}
We generate four independent graphs with $n=40$, using seeds 101, 202, 303, and 404. Node angles are independent uniform draws on $[-\pi,\pi)$, and unordered edges are conditionally independent Bernoulli variables with
\begin{equation}
p_{ij}^{\star}=\sigma\{-2.8+1.5\cos(t_i-t_j)\},\qquad i<j.
\end{equation}
The adjacency is symmetrized and its diagonal set to zero. Each graph is evaluated under $K=4$ and $K=8$ known groups. Node $i$, indexed from zero, belongs to group $i\bmod K$. Groups are balanced and independent of the simulated positions; the eight-group partition refines the four-group partition.

Every node responds, producing $Y=AB$. Both estimators receive only $Y$ and $B$, with no hidden edges, true angles, or pairwise covariates. The two values of $K$ reuse each graph, giving four independent graph replicates and a paired design, not eight independent graphs.

The circular estimator matches the generating geometry and fixes its strength at the generating value $1.5$. This supplies a structural advantage even though observed inputs are equal. The experiment examines a favorable geometric setting rather than averaging across unrelated network classes.

\subsection{Implemented estimators}
\subsubsection{Circular working posterior}
The fitted model estimates a global intercept $\alpha$ and 40 angles, using $p_{ij}=\sigma\{\alpha+1.5\cos(t_i-t_j)\}$. The prior is $\alpha\sim N(-2,2^2)$ with independent uniform angles. The likelihood is the Poisson working objective in~\eqref{eq:working}.

Each of two independently initialized chains runs 400 warmup sweeps followed by 800 retained sweeps. A sweep proposes a normal random walk for $\alpha$, then wrapped-normal random walks for every angle in random order. Metropolis acceptance uses the working likelihood and prior. Initial proposal scales are $0.2$ for the intercept and $0.7$ radians for angles. Every 50 warmup sweeps, each scale is multiplied by $\exp\{\operatorname{clip}(a-0.35,-0.3,0.3)\}$, where $a$ is the window acceptance rate, and clipped to $[0.02,3]$. Adaptation stops after warmup. No thinning is used.

The estimated matrix averages edge probabilities across all 1600 retained draws. Scalar diagnostics track the intercept, working log-likelihood, and aggregate mean-squared residual. The diagnostic screen uses classic split $\widehat R<1.05$ and an initial-positive-sequence effective sample size of at least 100 for each scalar. These summaries are invariant to geometric orientation.

\subsubsection{Regularized probability matrix}
The comparator solves
\begin{equation}
\min_{P\in\mathcal C}\ \tfrac12\|PB-Y\|_F^2+\|P\|_*,
\label{eq:nuclear}
\end{equation}
where $\|P\|_*$ is the nuclear norm and $\mathcal C$ is defined in~\eqref{eq:matrix}. The penalty coefficient is fixed at one for every run; hidden edges do not enter tuning. Its relative strength changes with the group design. This is a specified matrix comparator inspired by nuclear-norm regularization, not a replication of the full published Alidaee procedure.

A consensus ADMM solver splits $P$ into a nuclear-norm variable $L$ and a constrained variable $C$. With scaled duals $U,V$ and $\rho=2$, it alternates
\begin{align}
P&\leftarrow(YB^\top+\rho(L-U+C-V))(BB^\top+2\rho I)^{-1},\\
L&\leftarrow\operatorname{SVT}_{1/\rho}(P+U),\\
C&\leftarrow\Pi_{\mathcal C}(P+V),
\end{align}
followed by $U\leftarrow U+P-L$ and $V\leftarrow V+P-C$. The feasible-set projection symmetrizes, clips entries to $[0,1]$, and sets the diagonal to zero. All variables start at zero. Iteration stops when the recorded primal and dual Frobenius residuals are both below $10^{-6}$, with a cap of 10000 iterations. The returned estimate is the feasible matrix $C$.

\subsubsection{Density baseline}
The constant-density baseline uses
\begin{equation}
\widehat p=\frac{\sum_{i,k}y_{ik}}{n(n-1)}
\end{equation}
for every off-diagonal entry, with diagonal zero. Since groups partition all nodes and all respondents report, the numerator counts each realized edge twice. This baseline uses ARD alone.

\subsection{Evaluation}
We report three distinct metrics. AUC ranks unordered off-diagonal pairs against realized $A$, assigning half credit to ties. Probability RMSE compares fitted probabilities with $P^\star$ over the same pairs. ARD RMSE is $\|\widehat P B-Y\|_F/\sqrt{nK}$.

AUC measures reconstruction of the edges that generated the aggregate observations, not independent future-edge prediction. Probability RMSE measures the generating probabilities, and ARD RMSE measures in-sample aggregate fit. These targets can favor different estimators. Means and sample standard deviations are computed across four graph seeds separately for each $K$; the reported standard deviations are not standard errors.

\begin{table}[htbp]
\centering\small
\setlength{\tabcolsep}{3.5pt}
\begin{tabular}{rlrrr}
\hline
$K$ & Method & AUC & Probability RMSE & ARD RMSE \\
\hline
4 & Circular Bayes & 0.828 (0.052) & 0.071 (0.002) & 0.826 (0.084) \\
4 & Nuclear-norm comparator & 0.892 (0.031) & 0.123 (0.008) & 0.193 (0.008) \\
4 & Constant density & 0.500 (0.000) & 0.072 (0.002) & 0.875 (0.106) \\
8 & Circular Bayes & 0.908 (0.028) & 0.068 (0.003) & 0.566 (0.042) \\
8 & Nuclear-norm comparator & 0.966 (0.004) & 0.128 (0.010) & 0.204 (0.008) \\
8 & Constant density & 0.500 (0.000) & 0.072 (0.002) & 0.607 (0.062) \\
\hline
\end{tabular}

\caption{Matched-input reconstruction. Entries are mean (sample SD) across four graph replicates.}
\label{tab:rerun}
\end{table}

\subsection{Results}
\begin{figure}[htbp]
\centering
\includegraphics[width=\linewidth]{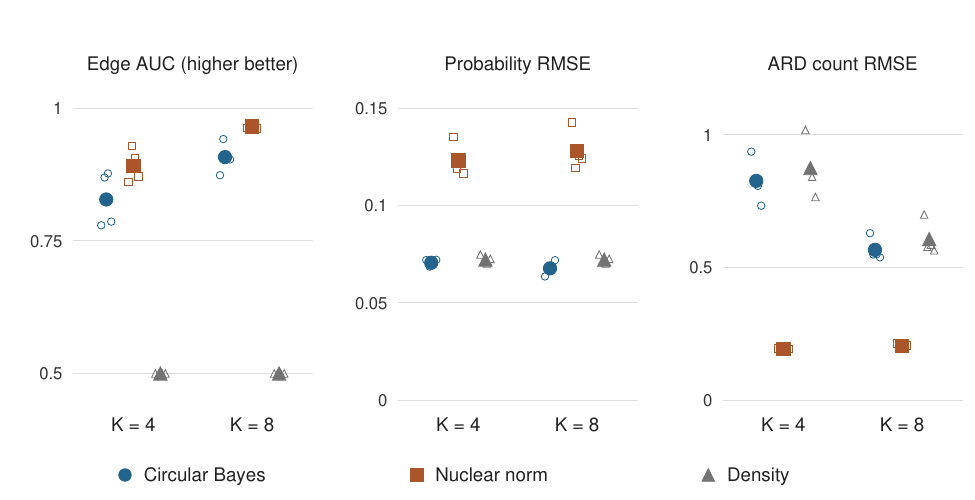}
\caption{Three reconstruction targets under matched inputs. Small open symbols show the four graph replicates; large filled symbols show their means. The same graphs are used at $K=4$ and $K=8$. Higher AUC and lower RMSE indicate better performance for their respective targets.}
\label{fig:metrics}
\end{figure}

\begin{figure}[htbp]
\centering
\includegraphics[width=.85\linewidth]{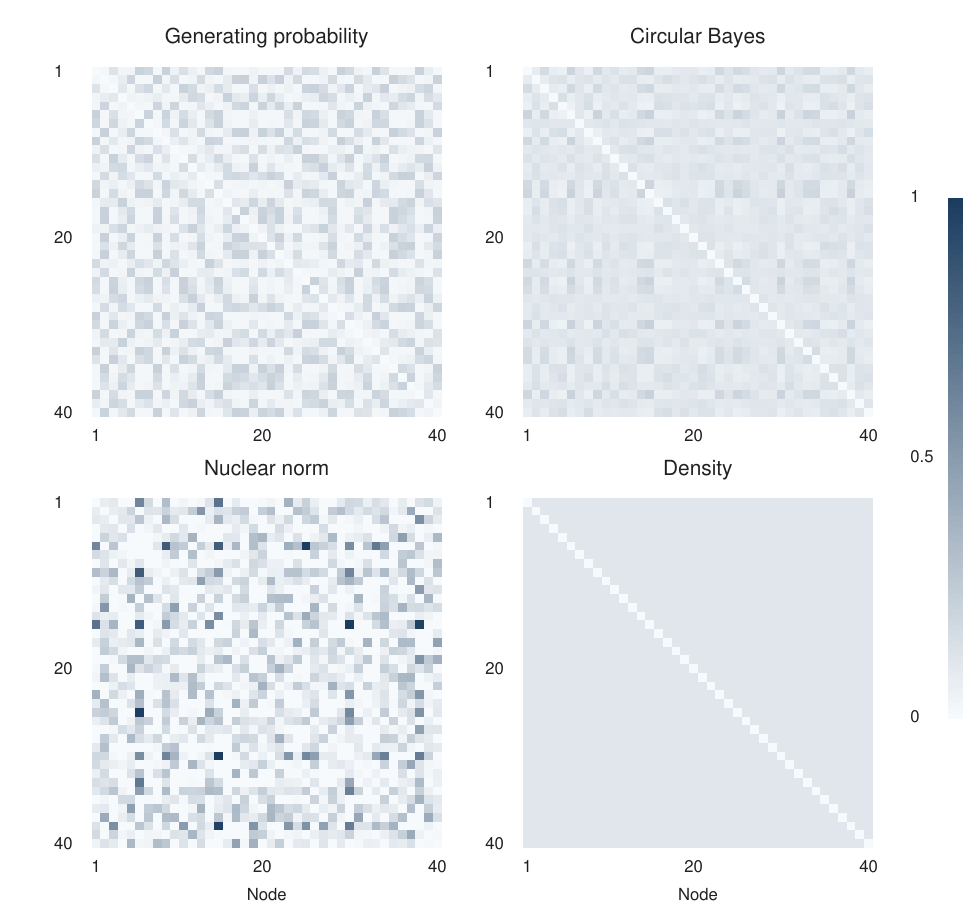}
\caption{Generating and estimated probabilities for seed 101, $n=40$, and $K=8$. All panels retain the original node order and use the same probability scale from zero to one, with zero diagonals. This is the first prescribed seed, not a selected best-performing replicate.}
\label{fig:matrices}
\end{figure}

Figure~\ref{fig:metrics} displays the individual graph replicates alongside their means. Figure~\ref{fig:matrices} compares the fitted probability matrices for the first prescribed seed under the eight-group design.

The nuclear-norm comparator has the highest realized-edge AUC: $0.892$ at $K=4$ and $0.966$ at $K=8$, compared with $0.828$ and $0.908$ for circular Bayes. It also fits the observed group counts more closely. The circular estimator has lower probability RMSE: $0.0706$ and $0.0677$, compared with $0.1232$ and $0.1280$ for the matrix comparator. The density baseline has probability RMSE $0.0722$ at both values of $K$.

Thus the ordering depends on the reconstruction target. A flexible fit can rank realized edges well while estimating their generating probabilities poorly. The circular estimator improves probability RMSE only modestly over constant density in these small graphs, despite its matched geometry. Increasing the number of groups improves AUC for both fitted methods, while the matrix comparator's probability RMSE does not improve in this run. These are paired finite-design comparisons, not evidence of a universal ordering.

All eight ADMM fits meet the specified residual tolerances. All eight MCMC scalar screens pass, with maximum split $\widehat R=1.018$ and minimum effective sample size approximately 244. These checks do not exclude mode trapping in latent configurations or establish posterior calibration.

Mean fitting times are approximately $2.5$ seconds for circular Bayes and range from $0.39$ to $0.65$ seconds for the matrix comparator. They refer to these $n=40$ implementations on the recorded local runtime with numerical thread counts fixed at one.

\subsection{Reproducibility}
The companion bundle contains NumPy estimation code, seeds, fitted matrices, inputs, evaluation truth, chain traces, diagnostics, and per-run metrics. Seven tests verify aggregation, estimator input boundaries, constraints, AUC ties, reproducibility, solver residuals, and saved results. Runtime details accompany the results.

\section{Applications and Design Implications}\label{chap:applications}
\subsection{From reconstruction to network summaries}
A fitted probability matrix can be used to compute expected degrees or generate networks for summary calculations. A centrality of a thresholded network, an average centrality over sampled networks, and centrality computed directly from a weighted matrix are different outputs. The reconstruction target should be chosen to match the scientific quantity of interest.

For interbank networks, a roster may contain institution types while aggregate reports summarize relationships with those types. Reconstructed connectivity is one input to risk analysis. Exposure sizes, balance sheets, and a contagion or loss model are also needed; centrality alone is not a default-loss estimate.

\subsection{Recommendation and contact networks}
Social platforms may disclose counts of contacts by group rather than individual links. Roster-based reconstruction can use known group memberships; survey-only reconstruction must integrate over unknown memberships. These settings should not be mixed when reporting performance.

For contact data, binary adjacency records whether contact exists, while weighted adjacency records frequency, duration, or another quantity. Replacing a Poisson distribution with a negative binomial distribution does not by itself specify how a weighted network generates observations.

\subsection{Trait design}
Group definitions determine which distinctions appear in the observations. Balanced groups may still supply similar restrictions. Adaptive designs could refine groups where uncertainty is greatest, while accounting for respondent burden and dependence introduced by adaptive selection.

Known group memberships do not reveal individual edges, but they do provide additional information relative to aggregate group sizes alone. Reporting this information boundary is part of describing the experiment, not a qualification added after the results.

\subsection{Computation and privacy}
Comparisons should specify the target objective, initialization, stopping criteria, retained draws, and hardware. Equal iteration counts need not imply comparable estimation accuracy. Mixing diagnostics for parameter summaries and predictive matrices also answer different questions.

A differential-privacy analysis specifies a neighboring-data relation, sensitivity, and release mechanism. These choices determine how disclosure protection interacts with the aggregate observation design. Distributed computation and formal privacy are separate design choices.

\section{Conclusion}
ARD reconstruction combines an observation map with a structural representation of unobserved edges. We organize latent-geometry and regularized estimation around the same roster-based counts, separating adjacency, probability, and parameter targets.

The likelihood and loss calculations connect estimation to the aggregate-mean map. The Poisson objective weights residuals by inverse means; the Huber objective clips residuals and changes the population estimating equation. Identification depends on the observation map and structural restrictions, whereas regularization selects a fit.

The matched-input experiment distinguishes three reconstruction targets. The matrix estimator better ranks realized edges and fits observed counts; the circular estimator better estimates generating probabilities, with a smaller improvement over the density baseline. Observation design, target selection, and computational calibration jointly determine the interpretation of a reconstruction.

\section*{Acknowledgements}
Hoc opus, professoribus honorandis, Lunae, Aliciae, Alici,
Lyrae, Verae, Nerae, familiae meae, atque uxori in aeternum
dilectae, reverenter consecratur.

In memoriam amoris veri, et in praesentia vocum novarum.

\appendix
\section{Likelihood and Loss Derivatives}\label{app:derivatives}
\subsection{Poisson objective}
Write \(d_{ik}=\nabla\mu_{ik}\) and \(D_{ik}=\nabla^2\mu_{ik}\). Differentiation gives
\begin{align}
\nabla\ell_W&=\sum_{i\in\mathcal R,k}(y_{ik}/\mu_{ik}-1)d_{ik},\\
\nabla^2\ell_W&=\sum_{i\in\mathcal R,k}
\left\{(y_{ik}/\mu_{ik}-1)D_{ik}
-\frac{y_{ik}}{\mu_{ik}^2}d_{ik}d_{ik}^\top\right\}.
\end{align}
At a correctly specified mean, \(\E y_{ik}=\mu_{ik}\), so
\begin{equation}
-\E\nabla^2\ell_W=\sum_{i\in\mathcal R,k}\frac{d_{ik}d_{ik}^\top}{\mu_{ik}}.
\end{equation}
Expected curvature uses marginal means; equality with score variance also depends on joint dependence.

\subsection{Huber objective}
Away from \(r_{ik}=\pm\xi\),
\begin{equation}
\nabla^2Q_n=\sum_{i\in\mathcal R,k}
\{\psi_\xi'(r_{ik})d_{ik}d_{ik}^\top-\psi_\xi(r_{ik})D_{ik}\}.
\end{equation}
The second term remains unless its expectation vanishes under the observation model. For a minimizing equation, curvature is the Hessian of the loss, not its negative. Let \(H_n\) be the integrated Hessian along the segment from \(\theta_\star\) to an interior stationary estimate \(\widehat\theta\), where the required differentiability holds. Then
\begin{equation}
H_n(\widehat\theta-\theta_\star)=-\nabla Q_n(\theta_\star).
\end{equation}
A norm bound requires lower singular-value control of \(H_n\) and a bound on the score. Neither is supplied by mean-zero measurement error alone.

\bibliographystyle{plainnat}
\bibliography{verified}
\clearpage
\section*{中文摘要}
\begin{center}
彙總關係資料下未知網絡連結重建\\
貝葉斯潛在曲面建模與懲罰迴歸法探討\\
曾彥暄
\end{center}
彙總型關係資料（aggregated relational data, ARD）記錄個人與屬性群體的連結數，而不揭露個別邊。本文以共同的觀測映射整理貝氏潛曲面模型與正則化網絡估計，區分實現的鄰接矩陣、條件邊機率與模型參數。研究設定為已知每個節點群體歸屬的名冊式 ARD；兩種估計方法使用相同的名冊及群體計數。

本文整理 Poisson 工作概似與 Huber 損失的導數，說明群體重疊、共享邊與報告誤差如何影響估計。幾何模型以潛在位置限制邊機率；正則化則在觀測相容的候選解之間取捨。識別取決於觀測映射與模型限制，不等同於數值最佳化解的唯一性。

我們以可重跑的合成實驗比較相同資訊下的重建結果，明訂資料生成方式、估計程序與評估目標。矩陣估計在實現邊排序與群體計數擬合上較佳；幾何估計在生成機率的均方誤差上較佳。這個共同架構將網絡重建整理為觀測設計、結構限制與計算方法共同決定的推論問題。

\noindent\textbf{關鍵詞：} 彙總型關係資料、網絡重建、潛在空間模型、正則化、識別
\end{document}